\documentclass[a4paper,11pt]{article}
\pdfoutput=1 
\usepackage{jheppub} 

\usepackage[T1]{fontenc} 

\usepackage{algorithm}
\usepackage{algpseudocode}

\usepackage{amsfonts,amssymb}
\usepackage{graphicx}
\usepackage{epsfig}
\usepackage{booktabs}
\usepackage{feynmp}
\usepackage{hyperref}
\usepackage{longtable}
\usepackage{amsmath}
\usepackage{ upgreek } 

\usepackage{color,xcolor,fancybox,epsf,rotating,colordvi}

\newcommand{\GeV}{~{\rm GeV}}
\newcommand{\TeV}{~\rm TeV}
\newcommand{\fbm}{{~\rm fb}^{-1}}

\title{\boldmath A Novel Scenario in the Semi-constrained NMSSM
}

\author{Kun Wang}
\author{and Jingya Zhu}
\affiliation{Center for Theoretical Physics, School of Physics and Technology, Wuhan University,\\ Wuhan 430072, China}

\emailAdd{wk2016@whu.edu.cn}
\emailAdd{zhujy@whu.edu.cn}

\abstract{
In this work, we develop a novel efficient scan method, combining the Heuristically Search (HS) and the Generative Adversarial Network (GAN), where the HS can shift marginal samples to perfect samples, and the GAN can generate a huge amount of recommended samples from noise in a short time.
With this efficient method, we find a new scenario in the semi-constrained Next-to Minimal Supersymmetric Standard Model (scNMSSM), or NMSSM with non-universal Higgs masses.
In this scenario,
(i) Both muon g-2 and right relic density can be satisfied, along with the high mass bound of gluino, etc. As far as we know, that had not been realized in the scNMSSM before this work.
(ii) With the right relic density, the lightest neutralinos are singlino-dominated, and can be as light as 0-12 GeV.
(iii) The future direct detections XENONnT and LUX-ZEPLIN (LZ-7 2T) can give strong constraints to this scenario.
(iv) The current indirect constraints to Higgs invisible decay $h_2\to \tilde{\chi}^0_1 \tilde{\chi}^0_1$ are weak, but the direct detection of Higgs invisible decay at the future HL-LHC may cover half of the samples, and that of the CEPC may cover most.
(v) The branching ratio of Higgs exotic decay $h_2\to h_1h_1, a_1a_1$ can be over 20 percent, while their contributions ($h_2\to4\tilde{\chi}_1^0$) to the invisible decay are very small.
}

\begin{document}
\maketitle
\flushbottom

\section{Introduction}
\label{sec:intro}

Higgs boson was discovered in 2012 \cite{Aad:2012tfa, Chatrchyan:2012xdj}, and its production rate in most channels coincides with the Standard Model (SM) prediction considering uncertainties \cite{Aad:2019mbh, CMS:2020gsy, Sopczak:2020vrs}.
While there are still chances for physics beyond the SM.
For example, for the branching ratio of Higgs boson invisible decay, the current excluding limits are only $26\%$ by ATLAS \cite{Aaboud:2019rtt} and $19\%$ by CMS \cite{Sirunyan:2018owy}, with all data at Run I and data of about $36\fbm$ at Run II.

Supersymmetry is a popular theory beyond the SM, which introduces a new internal symmetry between fermions and bosons.
Thus the large hierarchy problem can be solved, gauge coupling can be unified, and dark matter (DM) candidates can be provided, etc.
In the Minimal supersymmetric Standard Model (MSSM) with 7 free parameters at the electroweak scale, a SM-like 125 GeV Higgs can be afforded, but need large fine-tuning, and the branching ratio of Higgs boson invisible decay can be about $10\%$ at most \cite{Cao:2012fz, Cao:2012im, Cao:2012yn}.
The Next-to Minimal Supersymmetric Standard Model (NMSSM) with $\mathbb{Z}_3$ symmetry extends the MSSM by a complex singlet superfield $\hat{S}$, but introduces four more parameters.
In the graceful and simple model of fully-constrained NMSSM (cNMSSM), all Higgs and sfermion masses are assumed to be unified at the Grand Unified theoretical (GUT) scale, thus only four parameters at GUT scale are left free \cite{Kowalska:2012gs,Gunion:2012zd,Ellwanger:2010es,Panotopoulos:2010tw,LopezFogliani:2009np,Belanger:2008nt,Djouadi:2008uj,Ellwanger:2008ya,Hugonie:2007vd}.
These four or five parameters run according to the Renormalization Group Equations (RGEs), forming the spectrum of NMSSM at low energy scale.
While it was found that when considering all the constraints including muon g-2, the SM-like Higgs mass can not reach to 125 GeV in the cNMSSM\footnote{Notice that there are also some other ways to solve this problem, e.g., introducing right-handed neutrinos to the cNMSSM \cite{Cerdeno:2017sks}.} \cite{Gunion:2012zd, Kowalska:2012gs}, like these in the CMSSM, NUHM1 and NUHM2 \cite{Cao:2011sn, Ellis:2012aa, Bechtle:2015nua, Athron:2017qdc}.

In this work, we consider possible scenarios of Higgs invisible decay in the semi-constrained NMSSM (scNMSSM) \cite{Wang:2019biy, Wang:2018vxp, Ellwanger:2014dfa, Das:2013ta, Ellwanger:2018zxt, Ellwanger:2016sur}, which relaxing the Higgs masses at the GUT scale, and also called NMSSM with non-universal Higgs mass (NUHM).
As a simple and graceful SUSY model, the scNMSSM had attracted much attention.
In Refs.\cite{Ellwanger:2014dfa, Das:2013ta} the constraints of LHC and dark matter to scNMSSM was considered, while the muon g-2 was left aside;
in Ref.\cite{Wang:2018vxp} muon g-2 was satisfied, while dark matter relic density is not sufficient;
In Refs.\cite{Wang:2019biy,Ellwanger:2018zxt,Ellwanger:2016sur}, direct searches for the higgsino sector was considered;
In Ref.\cite{Nakamura:2015sya}, the extended model with right-handed neutrinos was considered.
In this work, we consider all constraints including muon g-2, and also try to get sufficient relic density.

In this work, to include constraints of muon g-2, etc., get sufficient relic density, and get as-large-as-possible branching ratio of Higgs invisible decay, we developed a novel efficient method to scan the parameter space, which consists of the Heuristically Search (HS) and the Generative Adversarial Network (GAN).
Note that in Refs.\cite{Ren:2017ymm, Abdughani:2019wuv}, the Machine Learning (ML) scan method has been used to explore the parameter space, and a scanning tool xBIT \cite{Staub:2019xhl} based on ML has been developed.
This ML scan is based on several classifiers, dealing with a Classification problem that each sample gets a probability of how much it could be a perfect sample.
This scan method also needs to generate samples in high-dimension space, which will cost very long time, (eg., when the dimension is 9 and each dimension has 100 grid, at least $100^9$ samples need to be generated).
On the contrary, we adopt a generative model, the Generative Adversarial Network (GAN) \cite{Goodfellow:2014upx}, which is a famous star in deep learning area and also gets much attention in high energy physics \cite{Paganini:2017dwg,Musella:2018rdi,Erdmann:2018jxd,DiSipio:2019imz,Otten:2019hhl,Lin:2019htn,Butter:2019cae,Bellagente:2019uyp,Butter:2019eyo}.
The GAN can directly generate samples with the similar distribution as the training samples. So with a well-trained GAN, we can get as many recommended samples as we want.
And with the HS we developed, we have a chance to shift some `bad' or `marginal' samples to `good' samples.
Combined with HS and GAN, we developed a novel method that can get a huge mount of surviving samples in a short time.
Then, we used this novel efficient method to study the parameter space of scNMSSM, under current constraints including LHC, B physics, muon g-2, and dark matter, etc.
We require our surviving samples to satisfy all these constraints, and part of them predict right relic density.
To study Higgs invisible decay, we require the LSP mass lighter than half of the SM-like Higgs mass ($m_{\tilde{\chi}_1^{0}} < m_{h_{\rm SM}}/2$), and the invisible branching ratio be as large as possible.
As can be seen from the following sections, this method is powerful in getting this novel scenario in the scNMSSM.

The rest of this paper is organized as follows.
In section 2, we briefly introduce the Higgs and electroweakino sectors of the scNMSSM, and our search strategy consisting of HS and GAN.
In section 3, we describe the detail of our scan process and then discuss the Higgs invisible decay and light dark matter in the scNMSSM.
Finally, we draw our conclusions in section 4.

\section{The semi-constrained NMSSM and the search strategy}
The NMSSM extends the MSSM particle content by adding a singlet superfield $\hat{S}$, which provides an effective $\mu$-term.
The superpotential of the $\mathbb{Z}_3$-invariant NMSSM is
\begin{equation}
    W_{\rm NMSSM} = y_u \hat{Q} \cdot \hat{H_u} \hat{u}^c + y_d \hat{Q} \cdot \hat{H_d} \hat{d}^c+ y_u \hat{L} \cdot \hat{H_d} \hat{e}^c + \lambda \hat{S} \hat{H}_u \cdot \hat{H}_d + \frac{\kappa}{3} \hat{S} ^3
\end{equation}
where the hats are used for superfields, $y_{u,d,e}$ stand for corresponding Yukawa couplings, and $\lambda$, $\kappa$ are dimensionless coupling constants.
When the singlet superfield $\hat{S}$ gets a vacuum expectation value (VEV), $\left \langle S \right \rangle=v_s$,
a effective $\mu$-term is generated dynamically from the term $\lambda \hat{S} \hat{H}_u \cdot \hat{H}_d$, with
\begin{equation}
    \mu_{\rm eff}=\lambda v_s \, .
\end{equation}
For convenience, in the following we refer to $\mu_{\rm eff}$ as $\mu$.
And the VEVs of the two doublet Higgs superfields $\hat{H}_u$ and $\hat{H}_d$ are $v_u$ and $v_d$ respectively, where $v_u^2 + v_d^2 =v^2 = (174 \GeV)^2$.

The soft SUSY breaking terms in the NMSSM are only different from the MSSM in several terms:
\begin{equation}
\label{eq:soft}
-\mathcal{L}_{\rm NMSSM}^{\rm soft} = -\mathcal{L}_{\rm MSSM}^{\rm soft}|_{\mu=0} + {m}_S^2 |S|^2
+ \lambda A_\lambda S H_u \cdot H_d + \frac{1}{3} \kappa A_\kappa S^3 + h.c. \,,
\end{equation}
where the $S$, $H_u$ and $H_d$ are the scalar components of the superfields respectively,
the ${m}_S^2$ is the soft SUSY breaking mass for single field $S$,
and the trilinear coupling constants $A_\lambda$ and $A_\kappa$ have mass dimension.

Unlike that in the CNMSSM or CMSSM, in the scNMSSM the Higgs sector is assumed to be non-universal at the GUT scale.
Then, at the GUT scale, the Higgs soft mass $m^2_{H_u}$,$m^2_{H_d}$ and $m^2_{S}$ are allowed to be different from $M^2_0+\mu^2$, and the trilinear couplings $A_\lambda$, $A_\kappa$ can be different from $A_0$.
Hence, in the scNMSSM, the complete parameter sector can be usually chosen as
\begin{equation}
\label{eq:parameter}
\lambda,\,\, \kappa,\,\, \tan\!\beta \!\!=\!\! \frac{v_u}{v_d},\,\, \mu,\,\, A_\lambda,\,\, A_\kappa,\,\, A_0,\,\, M_{1/2}, \,\,M_0  \,
\end{equation}
at the GUT scale.
While the parameters at low energy scale can be calculated in the RGEs running from these GUT-scale parameters.

\subsection{The Higgs and electroweakinos sector of the scNMSSM}

When the electroweak symmetry broken, the scalar component of superfields $\hat{H}_u$ , $\hat{H}_d$ and $\hat{S}$ can be written as
\begin{equation}
\label{eq:higgsfields}
H_u=\left(
\begin{array}{c}
H^+_u \\
v_u+\frac{ H^R_u + i H^I_u}{ \sqrt{2} }
\end{array} \right)\,, \quad
H_d=\left(
\begin{array}{c}
v_d+\frac{H^R_d + i H^I_d }{ \sqrt{2} }\\
H^-_d
\end{array} \right)\,, \quad
S = v_s + \frac{S^R+i S^I}{ \sqrt{2} } \,,
\end{equation}
where $H^R_u$, $H^R_d$, and $S^R$ are CP-even component fields, $H^I_u$, $H^I_u$,
and $S^I$ are the CP-odd component fields,
and the $H^+_u$ and $H^-_d$ are charged component fields.
In practice, it is convenient to rotate the fields as
\begin{eqnarray}
    H_1 &=& \cos \beta H_u
    + \varepsilon  \sin \beta    H_d^* =
    \begin{pmatrix} H^+ \\ \frac{ S_1 + i P_1}{ \sqrt{2} } \end{pmatrix} \\
    H_2 &=& \sin \beta H_u
    - \varepsilon  \cos \beta    H_d^* =
    \begin{pmatrix} G^+ \\ v + \frac{ S_2 + i G^0}{ \sqrt{2} } \end{pmatrix}\\
    H_3 &=& S = v_s + \frac{S_3+i P_2}{ \sqrt{2} }
\end{eqnarray}
where
$ \varepsilon = \left( \begin{smallmatrix}
             0 & 1 \\ -1 & 0
 \end{smallmatrix}  \right) $,
and $H_2$, $H_1$ and $H_3$ are the SM Higgs doublet, new doublet and singlet respectively.

In the basis $(S_1, S_2, S_3)$, the CP-even Higgs boson mass matrix $M_{S}^2$ is given by \cite{Miller:2003ay, Carena:2015moc}
\begin{align}
M_{S,1 1}^2 &=  M_A^2 + \left(m_Z^2- \lambda^2 v^2 \right)\sin^2 2 \beta \;+\Delta M_{S,11}^2 \,, \\
M_{S,2 2}^2 &=  m_Z^2 \cos^2 2 \beta + \lambda^2 v^2 \sin^2 2 \beta \;+\Delta M_{S,22}^2 \,,  \\
M_{S,1 2}^2 &=  - \frac{1}{2}\left(m_Z^2- \lambda^2 v^2\right) \sin 4 \beta \;+\Delta M_{S,12}^2 \,, \\
\label{ms33} M_{S,3 3}^2 &= \frac{1}{4} \lambda^2 v^2 \left(\frac{M_A}{\mu / \sin 2 \beta }\right)^2 + \kappa v_s A_\kappa +4(\kappa v_s)^2- \frac{1}{2} \lambda \kappa v^2 \sin 2 \beta \,,\\
M_{S,1 3}^2 &=  -\left( \frac{M_A^2}{2 \mu /\sin 2 \beta } + \kappa v_s \right) \lambda v \cos 2 \beta \,,  \\
M_{S,2 3}^2 &=  2 \lambda \mu v \left[1-\left( \frac{M_A}{2\mu / \sin 2 \beta} \right)^2 - \frac{\kappa}{2 \lambda } \sin 2 \beta  \right] \,,
\end{align}
where $M_A$ is the mass scale of new doublet with
\begin{equation}
    M_A^2 =   \frac{2\mu (A_\lambda + \kappa v_s)}{\sin 2\beta} \, ,
\end{equation}
and $\Delta M_{S,11}^2$,  $\Delta M_{S,22}^2$ and $\Delta M_{S,12}^2$ are the important corrections at loop level.
The first-order contributions by stop loops are given by \cite{Carena:2015moc}
\begin{eqnarray}
  \Delta M_{S,11}^2 &=& \frac{3v^2 y_t^4 \sin^2\!{2\beta}}{32\pi^2} \left[\ln \left(\frac{M_S^2}{m_t^2}\right) +\frac{X_t Y_t}{M_S^2} \left(1-\frac{X_t Y_t}{12M_S^2}\right) \right] \,, \\
  \Delta M_{S,22}^2 &=& \frac{3v^2 y_t^4 \sin^4\!\beta}{8\pi^2} \left[\ln \left(\frac{M_S^2}{m_t^2}\right) +\frac{X_t^2}{M_S^2}\left(1-\frac{X_t^2}{12M_S^2}\right) \right] \,, \\
  \Delta M_{S,12}^2 &=& \frac{3v^2 y_t^4 \sin^2\!\beta \sin\!{2\beta}}{16\pi^2} \left[\ln \left(\frac{M_S^2}{m_t^2}\right) +\frac{X_t(X_t+Y_t)}{2M_S^2} -\frac{X_t^3Y_t}{12M_S^2} \right] \,,
\end{eqnarray}
where $X_t = A_t - \mu/\tan\!\beta$, $Y_t = A_t + \mu \tan\!\beta$, $M_S=\sqrt{m_{\tilde{t}_1}m_{\tilde{t}_2}}$ is the geometric average of the two stop masses and $A_t$ is the trilinear parameter associated with the Yukawa coupling of top quark $y_t=m_t/v$.
To have the SM-like Higgs at about $125\GeV$, with $\tan\beta\gg1$ and $\lambda\ll 1$ the loop correction $\Delta M_{S,22}$ need to be about $(86\GeV)^2$, which means heavy stops ($M_S \!\thicksim\! 10\TeV$), or large stop mixing parameter $A_t$.

In the basis $(P_1, P_2)$, the CP-odd Higgs boson mass matrix $M_{P}^2$ is
\begin{align}
M_{P,1 1}^2 &=  M_A^2  \,, \\
M_{P,1 2}^2 &=  \lambda v (A_\lambda-2\kappa v_s) \,, \\
\label{mp22} M_{P,2 2}^2 &=  \lambda (A_\lambda+4\kappa v_s)  \frac{v_u v_d}{v_s}  - 3 \kappa v_s A_\kappa \,.
\end{align}

Three CP-even mass eigenstates $h_i~ (i=1,2,3)$ (ordered in mass) are mixed from $S_i ~(i=1,2,3)$, and two CP-odd mass eigenstates $a_i ~(i=1,2)$ (ordered in mass) are mixed from $P_i ~(i=1,2)$. The mixings are given by
\begin{equation}
\left( \begin{array}{c}
h_1  \\
h_2  \\
h_3
\end{array} \right)
= S_{ij}
\left( \begin{array}{c}
S_1  \\
S_2  \\
S_3
\end{array} \right) \,,
\qquad \quad
\left( \begin{array}{c}
a_1  \\
a_2
\end{array} \right)
= P_{ij}
\left( \begin{array}{c}
P_1  \\
P_2
\end{array} \right) \,,
\end{equation}
where the mixing matrix $S_{ij}$ and $P_{ij}$ can diagonalize the mass matrix $M_{S}^2$ and $M_{P}^2$ respectively.

The neutralino sector consists of five neutralinos.
In the gauge-eigenstate basis
$\psi^0 = (\tilde{B}, \tilde{W}^{3}, \tilde{H}_{d}^0, \tilde{H}_{u}^0, \tilde{S})$,
the neutralino mass matrix takes the form
\cite{Ellwanger:2009dp}
\begin{eqnarray}
\label{Neu-mass}
M_{\tilde{\chi}^{0}}= \left( \begin{array}{ccccc}
    M_{1}        & 0                 & -c_{\beta} s_W m_Z  &  s_{\beta} s_W m_Z  & 0 \\
    0            & M_{2}             &  c_{\beta} c_w m_z  & -s_{\beta} c_W m_Z  & 0 \\
-c_{\beta} s_W m_Z &  c_{\beta} c_w m_z  & 0               & -\mu             & -\lambda v_{d} \\
 s_{\beta} s_W m_Z & -s_{\beta} c_W m_Z  & -\mu            & 0                & -\lambda v_{u} \\
    0            & 0                 & -\lambda v_{d}      & -\lambda v_{u}   & 2\kappa v_s \\
\end{array} \right)
\end{eqnarray}
where $s_{\beta}=\sin\beta$, $c_{\beta}=\cos\beta$, $s_W=\sin\theta_W$, $c_W=\cos\theta_W$.
The mass eigenstates are denoted by $\tilde{\chi}^0_i$ $(i=1,2,3,4,5)$ ordered in mass.
Hereinafter $\tilde{\chi}^0_1$ is identified as the LSP.

Combining with Eq.\eqref{ms33}, Eq.\eqref{mp22}and Eq.\eqref{Neu-mass},we get a relation \cite{Das:2012rr,Ellwanger:2016sur} :
\begin{equation}
M_{\tilde{\chi}^{0},5 5 }^2  =  4 \kappa^2 v_s^2 = M_{S,3 3}^2 + \frac{1}{3} M_{P,2 2}^2 - \frac{4}{3} v_u v_d (\frac{\lambda^2 A_\lambda}{\mu}+\kappa) \,.
\end{equation}
If the LSP $\tilde{\chi}^0_{\rm 1}$ is highly singlino-dominated, $h_1$ and $a_1$ are singlet-dominated, and with a sizable $\tan\beta$, a not-too-large $A_\lambda$, and small $\lambda$ and $\kappa$, this equation can become:
\begin{equation}
    \label{relation}
    m_{\tilde{\chi}^0_{\rm 1}}^2 \approx m_{h_1}^2 + \frac{1}{3} m_{a_1}^2 \,.
\end{equation}

\subsection{The Heuristically Search (HS)}

Usually, We divide the samples into 2 categories according to whether or not the samples passed all constraints.
A sample that violated several constraints might be not good enough, but there is a chance that we can lead it to become a good sample.

In our case, we first leave aside the dark matter and muon g-2 constraints, only imposing other constraints in the \textsf{NMSSMTools}.
A sample that passes other constraints will get a score to evaluate how much it violates the dark matter and muon g-2 constraints, and we call it a `\textbf{marginal sample}'.

\begin{table}[h]
    \caption{The three types of samples: the bad, marginal and perfect samples.}
    \vspace{0.1cm}
    \label{3type}
    \centering
    \begin{tabular}{c|ccc}
    \hline
  & Type 1      & Type 2           & Type 3          \\ \hline
    \begin{tabular}[c]{@{}c@{}}The basic\\ constraints\end{tabular}               & $\times$         & $\checkmark$       & $\checkmark$             \\ \hline
    \begin{tabular}[c]{@{}c@{}}The dark matter and \\ muon g-2 constraints\end{tabular} &             ---         & $\times$           & $\checkmark$             \\ \hline
  & bad samples & marginal samples & perfect samples \\ \hline
  Score &   None       & >0 & =0 \\ \hline
    \end{tabular}

\end{table}

In Table \ref{3type}, we classify the samples into 3 types:  the bad, marginal and perfect samples.
For marginal and perfect samples, they will get a score to value how much they violate the constraints.
And we try to shift these marginal samples to satisfy the dark matter and muon g-2 constraints, becoming perfect samples.
The score function is given as:
\begin{equation}
\label{eq:score}
  f(\mathbf{X})=\sum_{i=1}^N  \max \left [ 1-\frac{O^i_{ \rm Theor. max}}{O^i_{\rm Exp. min}},0 \right ] + \max \left [ \frac{O^i_{\rm Theor. min}}{O^i_{\rm Exp. max}}-1,0 \right ] \, ,
\end{equation}
where $\mathbf{X}$ represent a marginal sample, $O^i$ means the $i$-th observable depending on $\mathbf{X}$, the $N$ means there are N kinds of different observables, the $O^i_{\rm Theor. min}$ and $O^i_{\rm Theor. max}$ are calculated with \textsf{NMSSMTools}, and the $O^i_{\rm Exp. min}$ and $O^i_{\rm Exp. max}$ are given by experimental results.
When the score is large, it means the marginal sample violates the experiments more; while when the score is zero, it means the marginal sample becomes a perfect sample, and satisfies all constraints very well, including dark matter and muon g-2 constraints.

In Algorithm~\ref{code:1}, we give the \textbf{Heuristically Search algorithm}, which can shift a marginal sample to a perfect sample satisfying all constraints.
With a marginal sample, $\mathbf{X}$, we search around it and try to find another marginal sample with a smaller score.
Then we repeat the process, until we meet a perfect sample whose score is zero, or get failed.

\begin{algorithm}[ht]
  \caption{The Heuristically Search (HS) with NMSSMTools}
  \label{code:1}
  \begin{algorithmic}[1]
    \Require
      A marginal sample, $\mathbf{X}$;
    \Ensure
      Find a perfect sample $\mathbf{X}$ passed all constraints, or failed;
    \label{code:Start}
    \State initial $step=0$ and $try=0$\;
    \State $score \gets f(\mathbf{X})$\;
    \While {$step < N_{max}$ \textbf{and} $try < T_{max}$ \textbf{and} $score \neq 0 $}
      \State get a new marginal sample $\mathbf{X'}$ around the $\mathbf{X}$ within radius $r$
      \State $score' \gets f(\mathbf{X'})$
       \If {$score' < score$}
            \State $\mathbf{X} \gets \mathbf{X'}$
            \State $score \gets score'$
            \State $step \gets step+1$
            \State $try \gets 0$
        \Else
            \State $try \gets try+1$
       \EndIf
    \State $//$ the search radius $r$ can be change with different $score$
    \EndWhile
    \If {$score = 0$}
        \State Succeed in getting a perfect sample $\mathbf{X}$
    \Else
        \State Failed
    \EndIf
    \label{code:End}
  \end{algorithmic}
\end{algorithm}

The search can be successful or get failed.
Most of the time in our case, the Heuristically Search can lead about $80\%$ (even over $94\%$) marginal samples to perfect samples.
Meanwhile, to avoid the program being trapped in a local minimum, we give it a chance to give up.
During the search, if the search step is larger than the maximum step $N_{\rm max}$ (we set it to 20), or the number of tries in one step is larger than the maximum number, $T_{\rm max}$ (we set it to 50), we stop the program and the search gets failed.

To get a new marginal sample $\mathbf{X'}$ around the $\mathbf{X}$, we can treat each component $\mathbf{x}_i$ ($i=1...9$) independently.
The simplest way is choosing samples around the $\mathbf{x}_i$ within radius $r_i$ with uniform distribution.
To improve the efficiency, the Gaussian distribution is adopted, since it has some chance to search samples far away and could jump out of the local minimums.
The Gaussian distribution function of $\mathbf{x'}_i$ is given as:
\begin{equation}
  f(\mathbf{x'}_i)=  \frac{1}{\sqrt{2 \pi}\sigma_i}exp\left[-\frac{ (x'_i-x_i)^2 }{2\sigma_i^2}\right] \, ,
\end{equation}
\begin{equation}
  \sigma_i= r_i  (x_{i,{\rm max}}-x_{i,{\rm min}}) \, ,
\end{equation}
where $r_i$ (we set it to $1/50$) is an important parameter and determines the search efficient.
Actually, $r_i$ can change with the score.
When the score is nearly zero, it means that a perfect sample is nearby, and then $r_i$ can change to a smaller one and vice versa.

\subsection{The Generative Adversarial Network (GAN)}

The Generative Adversarial Network (GAN) is a Generative model.
It can generate samples with similar distribution as the real data.
There are two neural networks in GAN.
One is the Generator $\mathbf{G}$, which can generate fake samples.
While the other is the Discriminator $\mathbf{D}$, which can classify the generated samples into real samples and the fake samples, so it is actually a binary classifier.

When the GAN is being trained, the Discriminator $\mathbf{D}$ tries to classify the generated samples into real and fake samples, meanwhile the Generator $\mathbf{G}$ tries to fool the Discriminator $\mathbf{D}$ and generate almost `real' samples.
After training, the Generator $\mathbf{G}$ and Discriminator $\mathbf{D}$ arrive at a Nash equilibrium.
Then we can use the Generator $\mathbf{G}$ to generate `real' samples as many as we need.
And these `real' samples actually have similar distribution as the real samples coming from the training dataset.

In this work, we use the Artificial Neural Networks to build the Generator $\mathbf{G}$ and the Discriminator $\mathbf{D}$.
We adopt a simple Neural Network with 3 hidden layers and each layer with 50 neurons, and the Activation Function is Leaky ReLU.
Furthermore, we train our GAN with Algorithm~\ref{code:2}.
In our case, we choose $k=3$, $n=1$, $m=20000$, and the training iterations as 2000, while for the Gradient descent we use Adadelta \cite{zeiler2012adadelta}.

\begin{algorithm}[ht]
  \caption{Training the Generative Adversarial Network (GAN)}
  \label{code:2}
  \begin{algorithmic}[1]
    \For{number of training iterations}
        \For{$k$ steps training the Discriminator}
            \State get $m$ perfect samples $\{\mathbf{X}^{(1)},\dots,\mathbf{X}^{(m)}\}$, from the training dataset;
            \State get $m$ noise samples, $\{\mathbf{z}^{(1)},\dots,\mathbf{z}^{(m)}\}$, generated by the Generator;
            \State update the Discriminator by descending its binary cross entropy
        \EndFor
        \For{$n$ steps training the Generator}
            \State get $m$ noise sample, $\{\mathbf{z}^{(1)},\dots,\mathbf{z}^{(m)}\}$, generated by the Generator
            \State update the Generator by descending its binary cross entropy
        \EndFor
    \EndFor
  \end{algorithmic}
\end{algorithm}

During the training, we require the Generator to learn the general distribution of the real data, but not try hard to find perfect hyperparameters, since we need the Generator to have more creativity.
As a complement, we combine GAN with the HS.
The Generator generates lots of samples, and some of them might be marginal samples, while the HS program will try to lead these marginal samples to perfect samples.

\section{Results and discussions}

To satisfy all the constraints including muon g-2, dark matter, Higgs data, gluino and other SUSY search results, and try to get right dark matter relic density and large Higgs invisible decay, we consider following parameter space in the scNMSSM:
\begin{eqnarray}
\label{eq:space}
0.1<\mu<0.2\TeV, \quad & 0<M_0<0.5\TeV, \quad & 0.5<M_{1/2}<2\TeV,
\nonumber\\
0.0<\lambda<0.7, \qquad & |\kappa|<0.7, \qquad & 1<\tan\beta <30,
\nonumber\\
|A_0|<10\TeV, \qquad & |A_\lambda|<10\TeV, \quad & |A_\kappa|<10\TeV .
\end{eqnarray}

\subsection{Scan with HS and GAN}

We developed the Heuristically Search program based on \textsf{NMSSMTools-5.5.2}
\cite{Ellwanger:2004xm,Ellwanger:2005dv,Ellwanger:2006rn,Das:2011dg}.
During the scan, we first require the samples satisfying the following other basic constraints:
\begin{itemize}
    \item Theoretical constraints of vacuum stability, and without Landau pole below $M_{\rm GUT}$ \cite{Ellwanger:2004xm,Ellwanger:2005dv,Ellwanger:2006rn}.
    \item The lower mass bounds of charginos and sleptons from the LEP:
\begin{equation}
    m_{\tilde{\uptau}} \ge 93.2 \GeV , ~~ m_{\tilde{\chi}_1^\pm} \ge 103.5 \GeV
\end{equation}
    \item Constraints from B physics, such as $B_s \to \mu^+ \mu^-$, $B_d \to \mu^+ \mu^-$, $b \to s \gamma$ and the mass differences $\Delta m_d$, $\Delta m_s$ \cite{Tanabashi:2018oca,Aaij:2012nna,Lees:2012xj,Lees:2012ym}
\begin{eqnarray}
    1.7 \times 10^{-9} <  Br(B_s \to \mu^+ \mu^-) < 4.5 \times 10^{-9} \\
    1.1 \times 10^{-10} < Br(B_d \to \mu^+ \mu^-) < 7.1 \times 10^{-10}  \\
    2.99 \times 10^{-4} < Br(b \to s \gamma) < 3.87 \times 10^{-4}
\end{eqnarray}

    \item An SM-liked Higgs boson exists with a mass between $123 \sim 127 \GeV$, and satisfies the global fit results with Higgs data at Run I and Run II of the LHC \cite{Aad:2019mbh, Sirunyan:2018koj, Khachatryan:2016vau}.
    \item To study Higgs invisible decay we require the mass of $\tilde{\chi}_1^{0}$ lighter than half of the SM-like Higgs,
        \begin{equation}
            m_{\tilde{\chi}^0_1} < \frac{1}{2} m_{h_{\rm SM}} \,.
        \end{equation}
\end{itemize}

Then for the marginal samples, we consider the constraints of dark matter and muon g-2, calculating the score in Eq.\eqref{eq:score} for each sample.
The upper and lower bounds of these observables are given in Table.\ref{5obs}.
\begin{table}[h]
    \centering
    \caption{The upper and lower bounds of the dark matter and muon g-2 observables.}
    \vspace{0.1cm}
    \label{5obs}
    \begin{tabular}{c|c|c}
    \hline
    & lower limit & upper limit             \\ \hline
    \begin{tabular}[c]{@{}c@{}}The DM relic \\ density $\Omega h^2$ \end{tabular}                               & None        & 0.131                 \\ \hline
    \begin{tabular}[c]{@{}c@{}}The spin-independent \\ DM-nucleon cross section\end{tabular} & None        & XENON1T                 \\ \hline
    \begin{tabular}[c]{@{}c@{}}The spin-dependent\\  DM-neutron cross section\end{tabular}   & None        & LUX and XENON1T         \\ \hline
    \begin{tabular}[c]{@{}c@{}}The spin-dependent \\ DM-proton cross section\end{tabular}    & None        & LUX, XENON1T and PICO-60 \\ \hline
    Muon g-2   $\delta a_\mu $                                                                               & $8.8\times 10^{-10}$         & $46\times 10^{-10}$                      \\ \hline
    \end{tabular}
\end{table}
The detail experimental constraints we consider in this work are list as following:
\begin{itemize}
\item The DM relic density $\Omega h^2$ from WMAP/Planck \cite{Hinshaw:2012aka, Ade:2013zuv, Tanabashi:2018oca}, we only take upper bound $\Omega h^2 \le 0.131$, considering there may be other sources of DM that contribute to $\Omega h^2$;
where the dark matter observables are calculated by \textsf{micrOMEGAs 5.0}
\cite{Belanger:2006is,Belanger:2008sj,Belanger:2010pz,Belanger:2013oya} inside \textsf{NMSSMTools}.
\item The spin-independent DM-nucleon cross section is constrained by XENON1T \cite{Aprile:2018dbl}, where we rescale the original values by $\Omega/\Omega_0$ with $\Omega_0 h^2=0.1187$;
\item The spin-dependent DM-nucleon cross section is constrained by LUX \cite{Akerib:2016lao}, XENON1T \cite{Aprile:2019dbj} and PICO-60 \cite{Amole:2019fdf}, where we also rescale the original values by $\Omega/\Omega_0$;
\item The muon anomalous magnetic moment (muon g-2) is constrained at $2\sigma$ level including all errors.
The difference between experimental result and SM theoretical value, including the corresponding error is given by \cite{Bennett:2006fi,Czarnecki:2002nt,Heinemeyer:2004yq,Bijnens:2007pz,Jegerlehner:2007xe}
\begin{equation}
    \delta a_\mu     \equiv a_\mu^{\rm ex} - a_\mu^{\rm SM} = (27.4 \pm 9.3) \times 10^{-10}
\end{equation}
where $a_\mu^{\rm SM}$ contains no Higgs contribution, since we consider a SM-like Higgs in SUSY contribution to $\delta a_\mu $.
We also consider the theoretical error of  SUSY contribution, which is about $1.5\times 10^{-10}$.
Thus at $2\sigma$ level, the central value of SUSY (including Higgses) contribution to muon g-2, $\delta a_\mu$, can be $5.8\!\sim\!49.0 \times 10^{-10}$.
\end{itemize}

If a sample satisfies the basic constraints (not including DM and muon g-2), it will get a score as a marginal or perfect sample; otherwise, it will be discarded.
Then with the HS program, we did our first scan.
We randomly searched for marginal samples in the parameter space, and then used the HS program changing them into perfect samples.
In the first search, we got about 10k perfect samples in 24 hours $\footnote{We used 40 threads parallel runing on Intel(R) Xeon(R) CPU E7-4830 v3 @ 2.10GHz.}$.
In fact if we changed the random scan into a multi-path Markov Chain Monte Carlo (MCMC) scan, the scan would be more efficient.

\begin{figure}[!tbp]
\centering
\includegraphics[width=.50\textwidth]{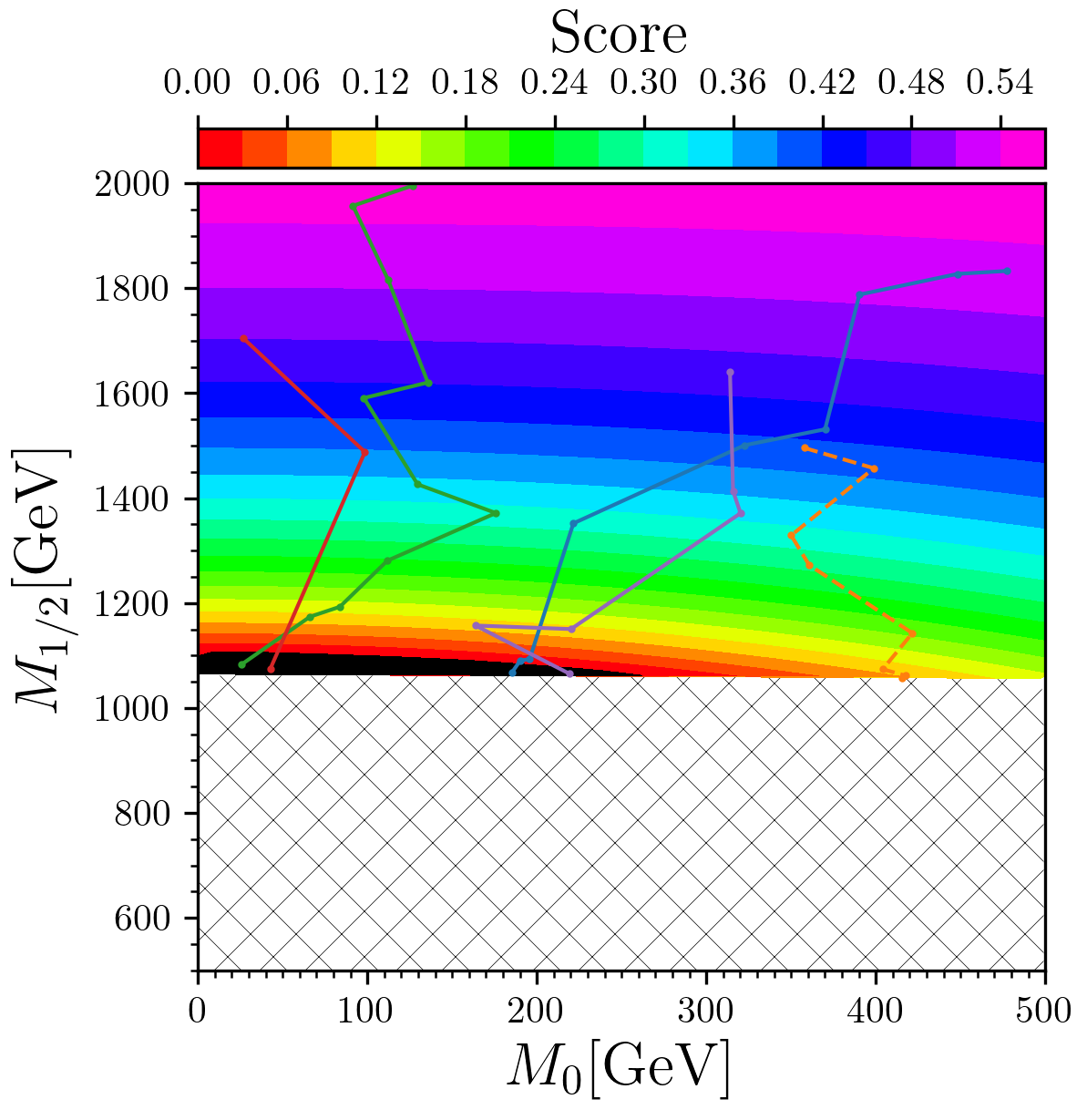}
\vspace{-0.5cm}
\caption{
The samples with seven parameters fixed ($\lambda=0.278$, $\kappa=-0.0577$, $\tan \beta = 17$, $\mu=162 \GeV$, $A_0=-1924 \GeV$, $A_\lambda=2756 \GeV$, and $A_\kappa=589 \GeV$) are projected in the $M_0$ versus $M_{1/2}$ plane.
The colored area indicate where the samples are marginal samples, and the colors indicate their score.
The black area indicate where the samples are perfect samples with zero score.
While the grid area indicate this piece of parameter space is excluded by the basic constraints.
The four solid lines indicate the marginal samples are led to become perfect samples, whereas the dash line indicates that the marginal sample is failed to shift to a perfect sample.
}
\label{fig1}
\end{figure}


In Fig.\ref{fig1}, we show the score of marginal samples in the $M_0$ versus $M_{1/2}$ plane.
Notice that if the score equal to zero, the marginal sample is also a perfect sample.
We can see that the area of marginal samples (colored range) is much larger than the perfect samples (black range) which get a zero score (satisfying all above constraints, including the DM and muon g-2).
Besides we also show five tries, that the HS program tries to shift marginal samples to perfect samples, where four get success (solid lines) and one gets failure (dashed line).
As the successful tries shown, the Heuristically Search usually needs less than 10 steps to shift a marginal sample to a perfect sample.
In fact, many marginal samples need only several steps to change into perfect samples, while the direct search for perfect samples will waste much more time.
That is the reason why we developed the HS program.

After the first search, all of the 10k perfect samples are used as the training set for the GAN.
Then we trained the GAN according to Algorithm~\ref{code:2}.
With a well-trained GAN $\footnote{We used Pytorch v1.3 to develop the GAN, and training cost about 5 hours. CPU: I5 6600K, GPU: GTX 1660 super. }$, we can transform random noises to recommended samples that have similar distribution as the training data.
Then we can easily get millions of recommended samples from the GAN in a few seconds.

\begin{figure}[!tbp]
\centering
\includegraphics[width=1\textwidth]{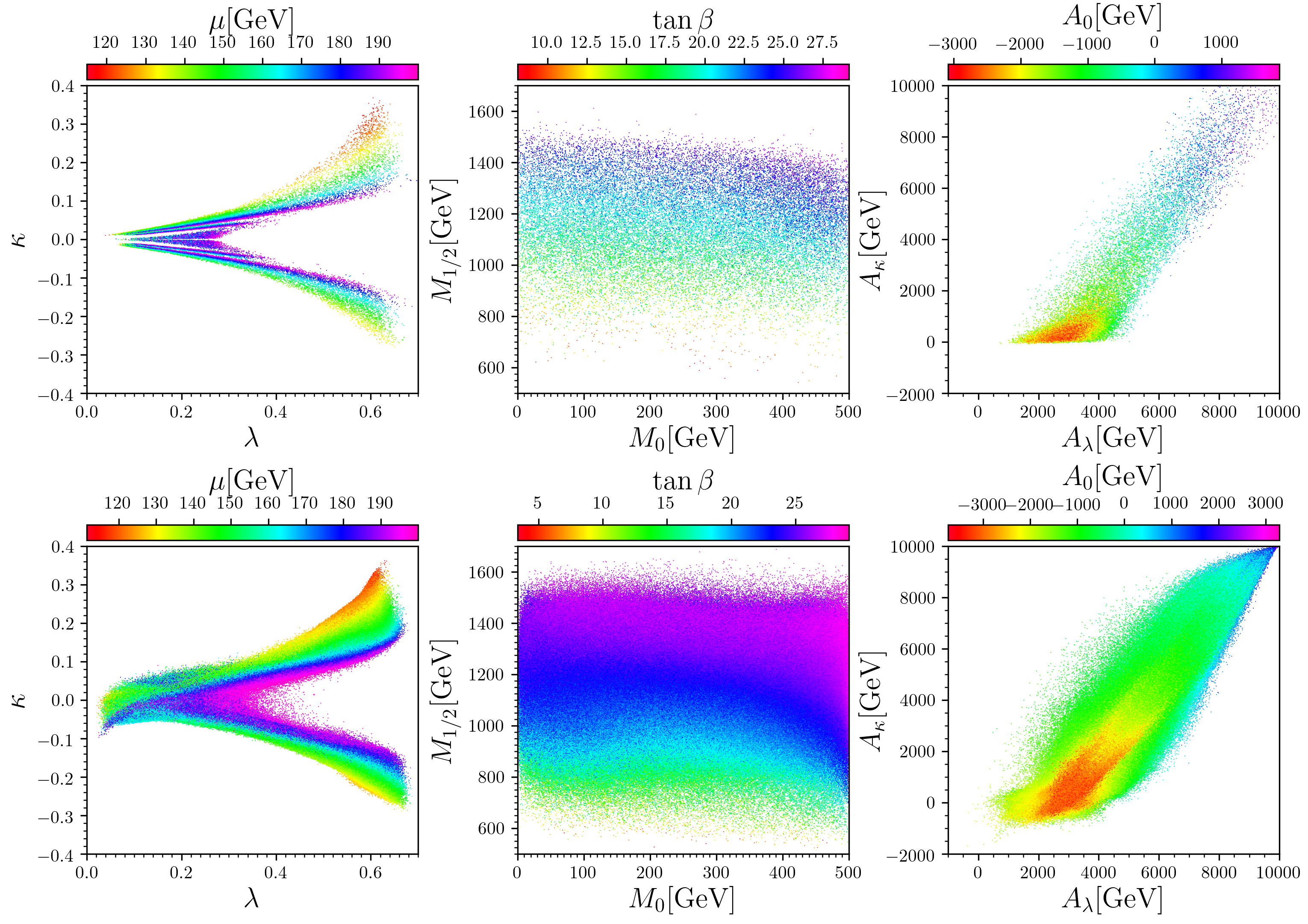}
\vspace{-0.8cm}
\caption{
The samples' distributions in the $\kappa$ versus $\lambda$ (left), $M_{1/2}$ versus $M_0$ (middle), $A_\kappa$ versus $A_\lambda$ (right) planes.
From left to the right, colors indicate the $\mu$, $\tan \beta$ and $A_0$, respectively.
\textbf{Upper Panel:} The training set for GAN, which are the original perfect samples generated in our first search.
\textbf{Lower Panel:} The recommended samples from GAN, which are generated by the Generator $\mathbf{G}$ in our well-trained GAN.
}
\label{fig2}
\end{figure}
In Fig.\ref{fig2}, we show the training set in the upper panels, and the recommended samples from GAN in the lower panels.
We can see that the GAN has already learned the general distribution of the perfect samples in the training set.
While the recommended samples from GAN (in the lower panels) have some creativity, which is not totally identical to the training set (in the upper panels).
The well-trained GAN can exploit the parameter space and recommend samples around the training samples, which is exactly what we need.

\begin{figure}[!tbp]
\centering
\includegraphics[width=1\textwidth]{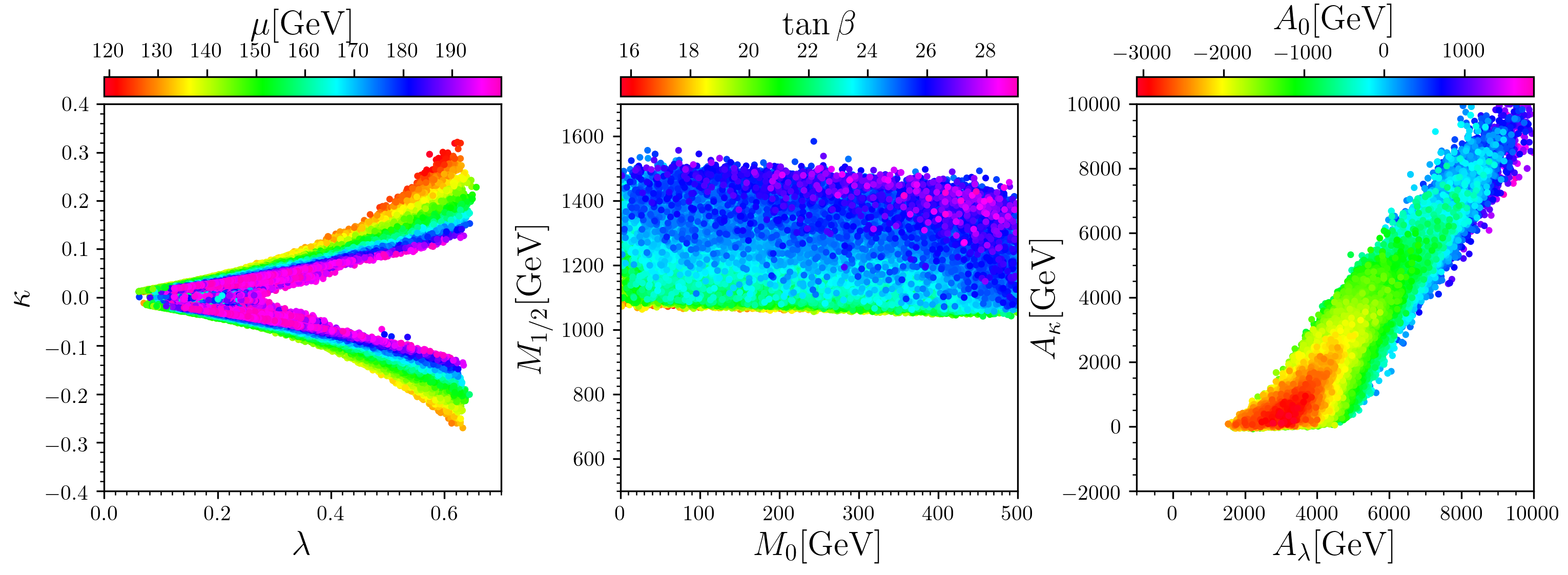}
\vspace{-0.8cm}
\caption{
The final surviving samples in the $\kappa$ versus $\lambda$ (left), $M_{1/2}$ versus $M_0$ (middle), $A_\kappa$ versus $A_\lambda$ (right) planes.
From left to the right, colors indicate the $\mu$, $\tan \beta$ and $A_0$, respectively.
}
\label{fig3}
\end{figure}

We used the trained GAN to generate 2000k recommended samples $\footnote{Less than 1 minute on the computer with CPU: I5 6600K, GPU: GTX 1660 super.}$, and passed these recommended samples to the HS program.
Then we got 280k perfect samples within 30 hours $\footnote{We used 40 threads parallel running on Intel(R) Xeon(R) CPU E7-4830 v3 @ 2.10GHz.}$, such a way is much faster than the traditional parameter scan.
At last, we impose the following additional constraints:
\begin{itemize}
    \item The upper limit of Higgs invisible decay, $19\%$, given by the CMS collaboration \cite{Sirunyan:2018owy}.
    \item The lower mass bound of colored sparticles, 
\begin{equation}
    \label{squarks and gluino}
    m_{\tilde{g}} > 2 {\TeV}, \, \,  m_{\tilde{t}_{1}} > 0.7 {\TeV}, \, \,  m_{\tilde{q}_{1,2}} > 2 {\TeV}.
\end{equation}
    \item The CMS constraints on charginos and neutralinos \cite{Sirunyan:2018ubx} implemented inside \textsf{NMSSMTools-5.5.2}.
    \item The SUSY search results implemented inside \textsf{SModelS v1.2.2} \cite{Kraml:2013mwa,Ambrogi:2017neo,Ambrogi:2018ujg,Dutta:2018ioj,Buckley:2013jua,Sjostrand:2006za,Sjostrand:2014zea} with official 1.2.2 database.
    \item The low- and high-mass resonances search results at the LEP, Tevatron and LHC, which are implemented inside \textsf{HiggsBounds-5.5.0} \cite{Bechtle:2015pma,Bechtle:2013wla,Bechtle:2013gu,Bechtle:2011sb,Bechtle:2008jh}.

\end{itemize}

Finally, after all the scans and constraints, we got about 88k surviving samples.
In Fig.\ref{fig3}, we show the nine free parameters of these surviving samples, and the coordinates are the same as those in Fig.\ref{fig2}.
We can see that all $M_{1/2}$ are larger than $1200 \GeV$. The reason is that we imposed the additional constraints, especially the high mass bound of gluino and the first-two-generation squarks at the LHC in Eq.\eqref{squarks and gluino}.

Comparing Fig.\ref{fig3} with the lower panels in Fig.\ref{fig2}, we can see that the recommended samples from GAN are changed to perfect samples by HS program.
While comparing Fig.\ref{fig3} with the upper plane in Fig.\ref{fig2}, we can see that the GAN has recommended many marginal samples that we need, and it does have some creativity to recommend samples around the training samples.
So, the combination of HS and GAN is very crucial.

\subsection{Light dark matter (DM) and Higgs invisible decay}
\begin{figure}[!tbp]
\centering
\includegraphics[width=1\textwidth]{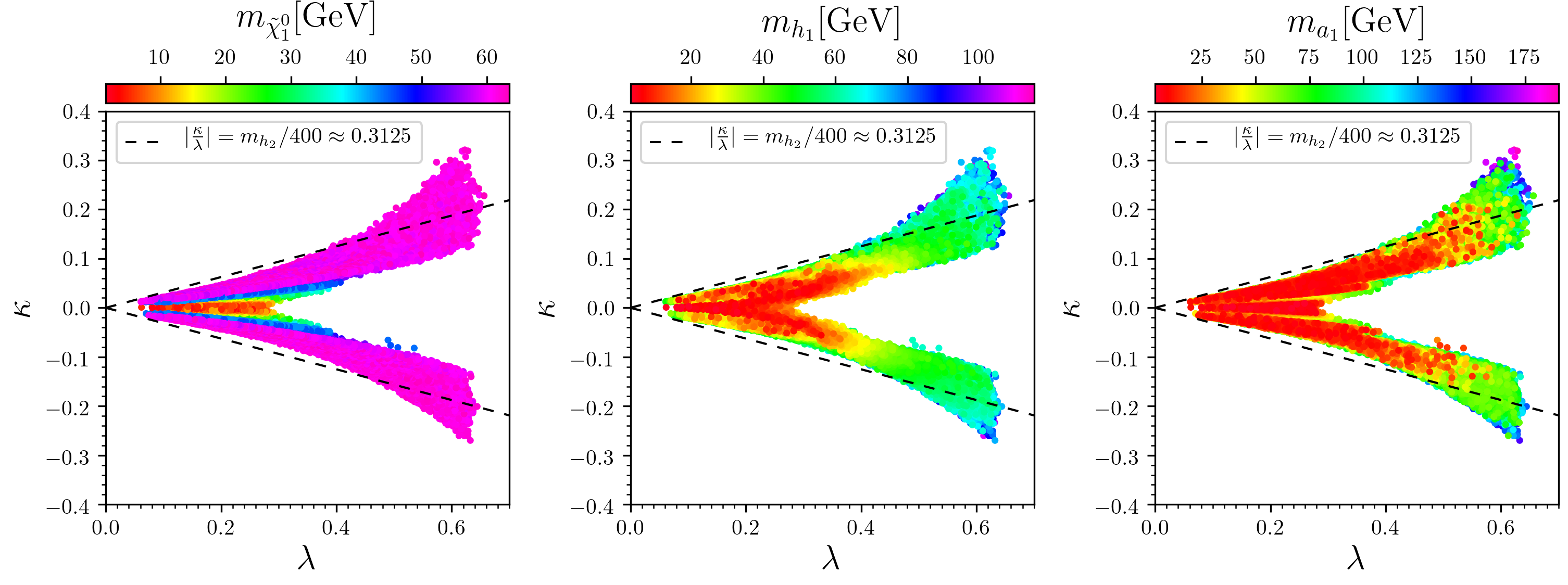}
\vspace{-0.8cm}
\caption{
The final surviving samples in the $\kappa$ versus $\lambda$ planes.
From left to the right, colors indicate the lightest neutralino (LSP) mass $m_{\tilde{\chi}_1^{0}}$, the lightest CP-even Higgs $h_1$ mass $m_{h_1}$ and the light CP-odd Higgs $a_1$ mass $m_{a_1}$, respectively.
The dash line is $\displaystyle \left|\frac{\kappa}{\lambda}\right|= 125/400 \approx 0.3125$.
In all these planes, samples with smaller mass are projected on top of the larger ones.
}
\label{fig4}
\end{figure}
In Fig.\ref{fig4} we show the final surviving samples in the plane of $\kappa$ vs $\lambda$, with colors indicate the masses of the lightest neutralino $\tilde{\chi}_1^{0}$, the lightest CP-even Higgs $h_1$  and the light CP-odd Higgs $a_1$ respectively.
For the surviving samples, we checked that the lightest CP-even Higgs $h_1$ are all highly singlet-dominated, and the next-to-lightest CP-even Higgs $h_2$ is the SM-like Higgs of 125 GeV.
Since we need the SM-like Higgs have a chance decaying to invisible $\tilde{\chi}_1^{0}$, the $\tilde{\chi}_1^{0}$ is lighter than $m_{h_2}/2$.
If the LSP $\tilde{\chi}_1^{0}$ is singlino-dominated, according Eq.\eqref{Neu-mass}, we should have
\begin{equation}
\label{s-LSP-mass}
  m_{\tilde{\chi}_1^{0}}=2\kappa v_s =2 \frac{\kappa}{\lambda} \mu \leq m_{h_2}/2 \, .
\end{equation}
Since we set the parameter $\mu$ from 100 to 200 GeV, we have
\begin{equation}
\label{kl}
 \left[  \frac{\kappa}{\lambda} \right]_{\rm max} \leq \left[ \frac{m_{h_2}}{4 \mu} \right]_{\rm min} =  \frac{m_{h_2}}{4 \times 100} \approx 0.3125 \, .
\end{equation}
Thus it is and we checked that the $\tilde{\chi}_1^{0}$ are singlino-dominated for samples between the two dash line.
We can also see that for the samples between the two dash lines, $h_1$ and $a_1$ are also possibly lighter than $m_{h_2}/2$.

\begin{figure}[!tbp]
\centering
\includegraphics[width=1\textwidth]{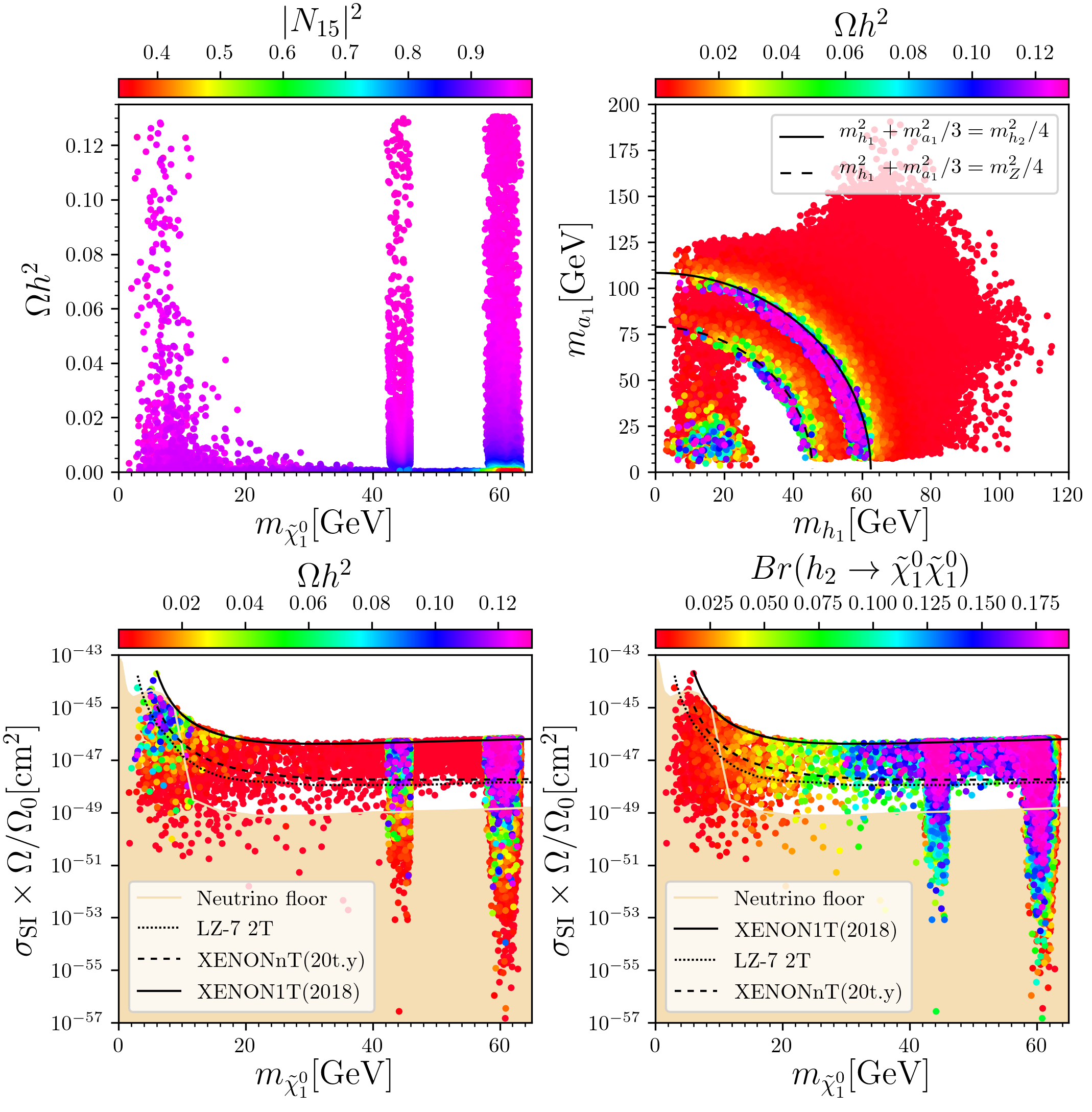}
\vspace{-0.8cm}
\caption{
\textbf{Upper Panel:} The surviving samples in the DM relic density $\Omega h^2$ versus the lightest neutralino (LSP) mass $m_{\tilde{\chi}_1^{0}}$ (left), and the CP-odd Higgs mass $m_{a_1}$ versus the CP-even Higgs mass $m_{h_1}$ (right) planes.
Colors indicate the singlino component $|N_{15}|^2$ in the $\tilde{\chi}_1^{0}$ (left), and the DM relic density $\Omega h^2$ (right) respectively.
In the right panel, the black solid and dashed curves indicate $m_{h_1}^2+ m_{a_1}^2/3=m_Z^2/4$ and $m_{h_1}^2+m_{a_1}^2/3=m_{h_2}^2/4$ respectively.
Samples with larger $|N_{15}|^2$ (left) or $\Omega h^2$ (right) are projected on top of the smaller ones.
\textbf{Lower Panel:} The surviving samples in the spin-independent DM-nucleon scattering cross section ($\sigma_{\rm SI} \times \Omega / \Omega_0 $) versus the LSP mass $m_{\tilde{\chi}_1^{0}}$ planes.
Colors indicate the DM relic density $\Omega h^2$ (left), and the Higgs invisible decay $Br(h_2 \to \tilde{\chi}_1^{0} \tilde{\chi}_1^{0})$ (right) respectively.
In these two panels, the black solid, dashed and dotted curves indicate the limits of spin-independent DM-nucleon cross section $\sigma_{\rm SI}$ by XENON1T 2018 \cite{Aprile:2018dbl}, the future detection sensitivity of XENONnT and LUX-ZEPLIN (LZ-7 2T), and the orange shaded region indicate the neutrino floor \cite{Billard:2013qya}.
Samples with larger $\Omega h^2$ (left) or $Br(h_2 \to \tilde{\chi}_1^{0} \tilde{\chi}_1^{0})$ (right) are projected on top of the smaller ones.
}
\label{fig5}
\end{figure}
\begin{figure}[!tbp]
\centering
\includegraphics[width=1\textwidth]{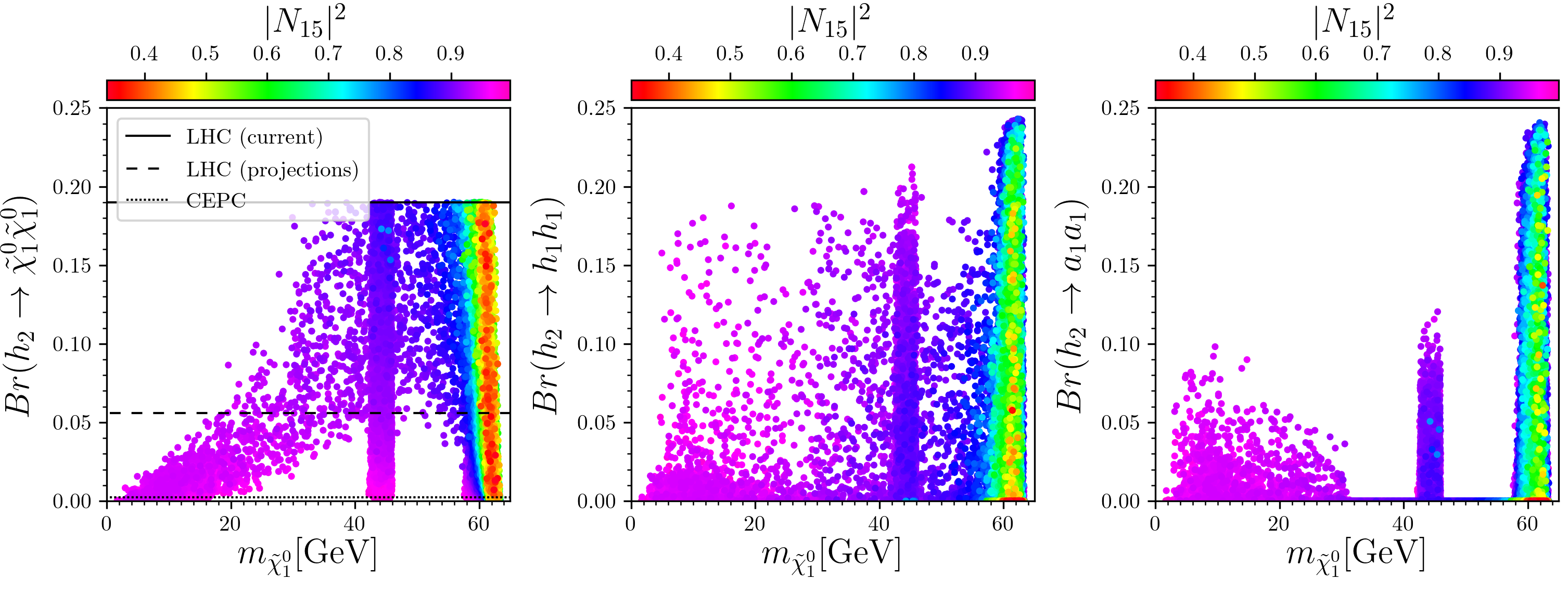}
\vspace{-0.8cm}
\caption{
The surviving samples in the Higgs invisible decay $Br(h_2\to \tilde{\chi}_1^{0}\tilde{\chi}_1^{0})$ (left), Higgs exotic $Br(h_2\to h_1 h_1)$ (middle) and $Br(h_2\to a_1 a_1)$ (right) versus the LSP mass $m_{\tilde{\chi}_1^{0}}$ planes respectively.
Colors indicate the singlino component $|N_{15}|^2$ in the LSP $\tilde{\chi}_1^{0}$.
In the left panel, the black dashed, dash-dotted and dotted lines indicate the Higgs invisible decay upper limit from the current LHC $19\%$ \cite{Sirunyan:2018owy}, future HL-LHC $5.6\%$ \cite{Liu:2016zki}, and CEPC $0.24\%$ \cite{Tan:2020fxk} respectively.
Samples with smaller $|N_{15}|^2$ are projected on top of the larger ones.
}
\label{fig6}
\end{figure}

In Fig.\ref{fig5} we show the properties of dark matter in the scNMSSM.
In the lower panels, the spin-independent dark matter and nucleon scattering cross section $\sigma_{\rm SI}$ have been rescaled by a ratio of $\Omega / \Omega_0 $, where the $\Omega_0$ is the right dark matter relic density with $\Omega_0 h^2 = 0.1187$.
As seen from these panels, the samples with right relic density can be divided into three cases:
\begin{itemize}
    \item \textbf{Case I:} $m_{\tilde{\chi}_1^{0}} \simeq m_{h_2}/2$
    \item \textbf{Case II:} $m_{\tilde{\chi}_1^{0}} \simeq m_{Z}/2$
    \item \textbf{Case III:} $m_{\tilde{\chi}_1^{0}} \lesssim 12 \GeV$
\end{itemize}
From Fig.\ref{fig5}, we can obtain the following observations:
\begin{itemize}
    \item From the upper left panel, the samples with right DM relic density are all with highly singlino-dominated $\tilde{\chi}_1^{0}$, where $|N_{15}|^2 \gtrsim 0.9$.
    \item From the upper right panel, there is a special relationship between the mass of $h_1$, $a_1$ and $\tilde{\chi}_1^{0}$.
    For the samples with right DM relic density in Case I and Case II, the LSP $\tilde{\chi}_1^{0}$ is highly singlino-dominated, and with small $\lambda$, $\kappa$ and a sizable $\tan\beta$.
    Combining with Eq.\eqref{relation}, we can see the two ellipse arcs:
    \begin{eqnarray}
    {\rm Case~I:} ~~~~~  m_{h_1}^2 + \frac{1}{3} m_{a_1}^2 \backsimeq m_{\tilde{\chi}^0_{\rm 1}}^2 \backsimeq \left(\frac{m_{h_2}}{2}\right)^2\\
    {\rm Case~II:} ~~~~~  m_{h_1}^2 + \frac{1}{3} m_{a_1}^2 \backsimeq m_{\tilde{\chi}^0_{\rm 1}}^2 \backsimeq \left(\frac{m_{Z}}{2}\right)^2
    \end{eqnarray}

    \item From the lower-left panel, most samples predict spin-independent DM-nucleon cross section $\sigma_{\rm SI}$ not far below the bound from XENON1T 2018, and can be covered by future LZ and XENONnT experiments.
    Thus these two future direct detections are crucial to check the parameter space of the scNMMSM.
    But there are still some samples that can escape from these future detections, and also can predict right relic density.
    Besides, there are also some samples below the neutrino floor, although most of them do not predict sufficient DM relic density.

    \item From the lower right panel, samples with large Higgs invisible decay branching ratio, $Br(h_2 \to \tilde{\chi}_1^{0} \tilde{\chi}_1^{0})>10\%$, have a sizable LSP mass, $m_{\tilde{\chi}^0_1}>30 \GeV$.
    This is because the small LSP mass, $m_{\tilde{\chi}^0_1}<30 \GeV$, always accompanying with a small $h_1$ and $a_1$ mass, which can be seen from the upper right panel of Fig.\ref{fig4}.
    Then the exotic decay channels $h_2 \to h_1 h_1$ and $h_2 \to a_1 a_1$ will open, which can be seen in Fig.\ref{fig6}.
    The Higgs invisible decay branching ratio $Br(h_2 \to \tilde{\chi}_1^{0} \tilde{\chi}_1^{0})$ become smaller.

    \item From the lower right panel, most samples which have large Higgs invisible decay branching ratio, $Br(h_2 \to \tilde{\chi}_1^{0} \tilde{\chi}_1^{0})>10\%$, could be covered by future LZ and XENONnT detections.
    But there are still some samples that can escape from these future experiments, and also can have large Higgs invisible decay branching ratio.
    And there are also some samples below the neutrino floor, some of them can have large Higgs invisible decay branching ratio $Br(h_2 \to \tilde{\chi}_1^{0} \tilde{\chi}_1^{0})>10\%$.
\end{itemize}

In Fig.\ref{fig6}, we show the decay information of the SM-like Higgs $h_2$.
From this figure, we can see that all of the branching ratios of $h_2\to \tilde{\chi}_1^{0}\tilde{\chi}_1^{0}, h_1h_1, a_1a_1$ can be at most about $20\%$.
While we checked that considering in addition that of $h_2$ decay to $4\tilde{\chi}_1^{0}$ though $a_1/h_1\to \tilde{\chi}_1^{0} \tilde{\chi}_1^{0}$,
which acquire $m_{\tilde{\chi}_1^{0}} < m_{h_2}/4 \simeq 31 \GeV $,
the branching ratio of Higgs invisible decay increase very little compared with only that $h_2$ decay to two $\tilde{\chi}_1^{0}$ though $h_2 \to \tilde{\chi}_1^{0} \tilde{\chi}_1^{0}$.
The upper limit of Higgs invisible decay branching ratio is about $19\%$ at Run II of the LHC, while the future detections for that can reach to $5.6\%$, $0.24\%$, $0.5\%$ and $0.26\%$ according to HL-LHC \cite{Liu:2016zki}, CEPC \cite{Tan:2020fxk}, FCC \cite{dEnterria:2017dac} and ILC \cite{Ishikawa:2019uda} respectively.

Considering the values of $|N_{15}|^2$, we can have the following observations from Fig.\ref{fig6}:
\begin{itemize}
    \item For most samples with higgsino-dominated LSP, $|N_{15}|^2 < 0.5$, the branching ratio $Br(h_2 \to \tilde{\chi}_1^{0} \tilde{\chi}_1^{0})$ can be sizeable, while the branching ratio $Br(h_2 \to h_1 h_1)$ and $Br(h_2 \to a_1 a_1)$ are both zero.
        The reason is that the higgsino-dominated LSPs are usually accompanied by a large mass of $h_1$ and $a_1$, as can be seen from the upper panels of Fig.\ref{fig5}, thus these two exotic decay channels are closed.
    \item For samples with $h_2/Z$-funnel dark matter, $m_{\tilde{\chi}_1^{0}}\simeq m_{Z,h_2}$, the branching ratio of Higgs boson invisible decay can be large or small depending on the parameter $\lambda$.
    \item For most samples with low-mass LSP, $m_{\tilde{\chi}_1^{0}}< 20\GeV$, the branching ratio of Higgs boson invisible decay is small and beyond the ability of HL-LHC, while the $Br(h_2 \to h_1 h_1)$ can be larger than the $Br(h_2 \to a_1 a_1)$.
    \item Though the detection of Higgs invisible decay, about half of the surviving samples can be covered at the future HL-LHC, while the future CEPC can cover most.
\end{itemize}

\begin{figure}[!tbp]
\centering
\includegraphics[width=1\textwidth]{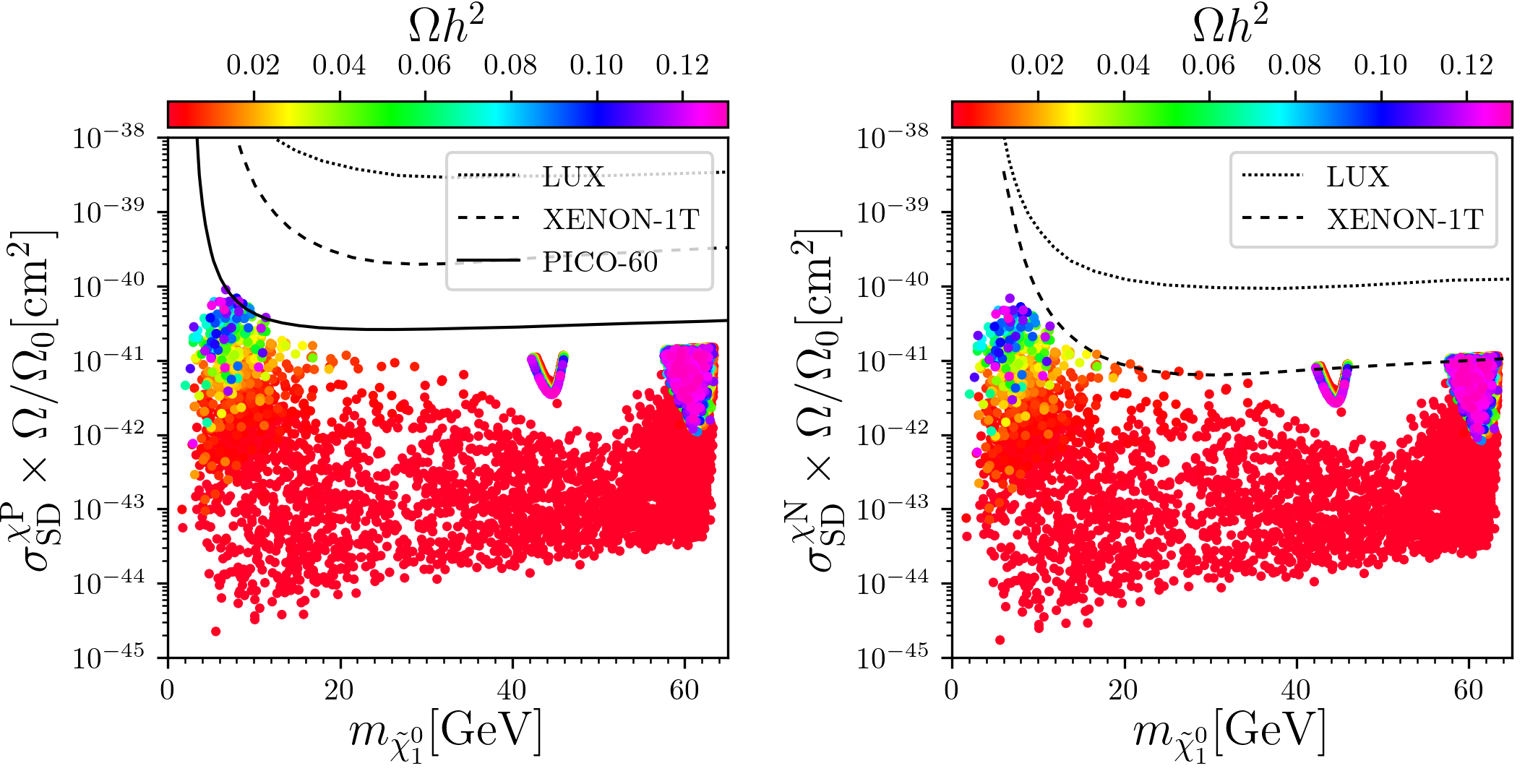}
\vspace{-0.8cm}
\caption{
The surviving samples in the spin-dependent cross section of DM-proton scattering $\sigma_{\rm SD}^{\rm \chi P}$ (left) and DM-neutron scattering $\sigma_{\rm SD}^{\rm \chi N}$ (right) versus the LSP mass $m_{\tilde{\chi}_1^{0}}$ planes respectively.
Colors indicate the LSP ratio in current dark matter $\Omega/\Omega_0$.
In the left panel, the black dotted, dashed and solid curves indicate the limits of spin-dependent DM-proton cross section $\sigma_{\rm SD}^{\rm \chi P}$ by LUX \cite{Akerib:2016lao}, XENON-1T \cite{Aprile:2019dbj} and PICO-60 \cite{Amole:2019fdf} respectively.
While in the right panel the black dotted and dashed curves indicate these of DM-neutron cross section $\sigma_{\rm SD}^{\rm \chi N}$ by LUX \cite{Akerib:2016lao}, and XENON-1T \cite{Aprile:2019dbj} respectively.
Samples with larger $\Omega h^2$ are projected on top of the smaller ones.
}
\label{fig7}
\end{figure}

\begin{figure}[!tbp]
\centering
\includegraphics[width=0.5\textwidth]{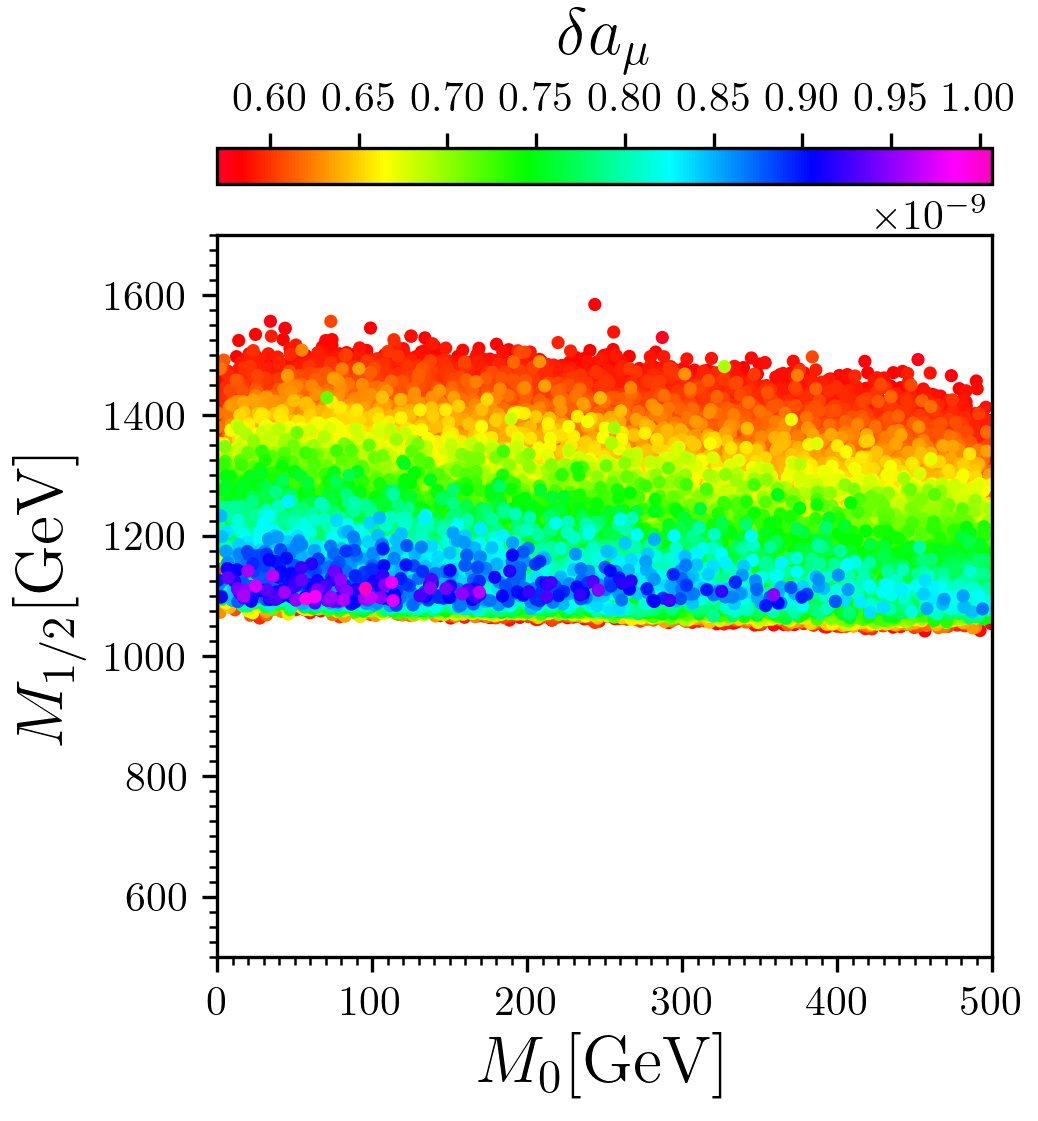}
\vspace{-0.6cm}
\caption{
The surviving samples in the parameter $M_{1/2}$ versus $M_0$ planes, with colors indicating $\delta a_\mu$, the central value of SUSY (including Higgses) contribution to muon g-2.
Samples with larger $\delta a_\mu$ are projected on top of the smaller ones.
}
\label{fig8}
\end{figure}

In addition, we list some discussions on other related topics in this scenario:
\begin{itemize}
  \item We had performed a work on the annihilating mechanisms of light dark matter in this scenario \cite{Wang:2020dtb}, where we found that all the samples have the LSP in funnel mechanisms. When the LSP is lighter than $20\GeV$, it is in $h_1$- or $a_1$-funnel mechanism, that is $2m_{\tilde{\chi}^0_1}\!\backsimeq\! m_{h_1}$ or $2m_{\tilde{\chi}^0_1}\!\backsimeq\! m_{a_1}$.
  \item \textsf{Higgsbounds} has been used to constrain heavy Higgs bosons. We also checked that the heavy bosons $h_3$ and $a_2$ are at $2.4\!\sim\!4.8\TeV$, and their branching ratios to $\tau$ pairs are $8\%$ at most.
      The masses are not covered in Ref.\cite{Aaboud:2017sjh}, and the production rates are much smaller than the upper limits in Ref.\cite{Sirunyan:2018zut}.
      Furthermore, we are ongoing a new work on the heavy Higgs bosons, especially on how to probe them at the future 100-TeV hadronic collider.
\item We again checked the spin-dependent cross sections, and show them in Fig.\ref{fig7}. As can be seen from it, both the DM-proton and DM-neutron cross sections satisfy the current constraints. When the LSP density $\Omega h^2$ is sufficient, the upper limit is satisfied directly; while when the LSP density $\Omega h^2$ is insufficient considering there may be other source of dark matter, the upper limit is satisfied by rescaling the cross section by a factor $\Omega/\Omega_0$, which is the ratio of LSP $\tilde{\chi}^0_1$ in current dark matter.
  \item We also checked muon g-2, and show $\delta a_\mu$, the central value of SUSY (including Higgses) contribution, in Fig.\ref{fig8}. When imposing the constraint, we also consider the error in SUSY-contribution calculation, which is about $1.5\times 10^{-10}$, thus all the samples can satisfy the experimental result at $2\sigma$ level.
      We also noticed that, the large $M_{1/2}$ values are caused by the high mass bounds of gluino and squarks in the first two generations, and this in return cause heavy wino-like chargino and bino-like neutralino, thus the SUSY contribution $\delta a_\mu$ cannot increase more.
\end{itemize}

\section{Conclusions}

In this work, we develop a novel scan method, combining the Heuristically Search (HS) and the Generative Adversarial Network (GAN).
The HS can shift marginal samples to perfect samples, and the GAN can generate recommended samples as many as we need from noise.

In our specific process, we first scan the parameter space randomly with $\textsf{NMSSMTools}$ under basic constraints, generating marginal samples;
then the HS try to shift the marginal samples to perfect samples satisfying in addition the dark matter and muon g-2 constraints;
with these randomly-generated perfect samples, the GAN is trained, and then generates a huge amount of recommended samples in a short time;
again the HS try to shift the recommended samples to perfect samples;
finally, we check the final perfect samples with additional constraints including these of sparticle searches, Higgs searches and Higgs invisible decay, getting the final surviving samples.

With this efficient method, we find a new scenario in the semi-constrained Next-to Minimal Supersymmetric Standard Model (scNMSSM), or NMSSM with non-universal Higgs masses.
In this scenario,
\begin{itemize}
  \item Both muon g-2 and right relic density can be satisfied, along with the high mass bound of gluino, etc. As far as we know, that had not been realized in the scNMSSM before this work.
  \item With the right relic density, the lightest neutralinos are singlino-dominated, and can be as light as 0-12 GeV.
  \item The future direct detections XENONnT and LUX-ZEPLIN (LZ-7 2T) can give strong constraints to this scenario.
  \item The current indirect constraints to Higgs invisible decay $h_2\to \tilde{\chi}^0_1 \tilde{\chi}^0_1$ are weak, but the direct detection of Higgs invisible decay at the future HL-LHC may cover half of the samples, and that of the CEPC may cover most.
  \item The branching ratio of Higgs exotic decay $h_2\to h_1h_1, a_1a_1$ can be over 20 percent, while their contributions ($h_2\to4\tilde{\chi}_1^0$) to the invisible decay are very small.
\end{itemize}


\acknowledgments

This work was supported by the National Natural Science Foundation of China (NNSFC)
under grant Nos. 11605123.


\bibliographystyle{JHEP}
\bibliography{ref}

\end{document}